\RequirePackage{tikz}
\documentclass{eptcs}
\providecommand{\event}{ADG 2023}

\usepackage{argoclp}
\usepackage{hyperref}
\usepackage{amsthm}

\newenvironment{figures}[4][H]
{\vspace{0.5em}~\refstepcounter{figure}
\begin{center}
\setlength{\unitlength}{#1 mm} {\small
\resizebox{#1\height}{!}{\includegraphics{#2}}
} \setlength{\unitlength}{1.0 mm}
\end{center}
\label{#4}
\centerline{{\bf Fig. \thefigure:} #3}
\par
\vspace{0.5em} \noindent }{}

\theoremstyle{definition}
\newtheorem{example1}{Example}[section]

\theoremstyle{definition}
\newtheorem{theorem1}{Example}[section]

\theoremstyle{lemma}
\newtheorem{lemma1}{Example}[section]

\title{Towards Automated Readable Proofs of\\ Ruler and Compass Constructions}

\author{Vesna Marinkovi\'c\institute{Faculty of Mathematics, University of Belgrade}\email{vesna.marinkovic@matf.bg.ac.rs} \and Tijana \v{S}ukilovi\'c\institute{Faculty of Mathematics, University of Belgrade}\email{tijana.sukilovic@matf.bg.ac.rs} \and Filip Mari\'c\institute{Faculty of Mathematics, University of Belgrade}\email{filip.maric@matf.bg.ac.rs}}

\def\titlerunning{Towards Automated Readable Proofs of Ruler and Compass Constructions}
\def\authorrunning{V. Marinkovi\' c, T. \v Sukilovi\' c \& F. Mari\' c}

\begin{document}
\maketitle

\begin{abstract}
  Although there are several systems that successfully generate
  construction steps for ruler and compass construction
  problems, none of them provides readable synthetic correctness proofs for generated
  constructions. In the present work, we demonstrate how our triangle
  construction solver ArgoTriCS can cooperate with automated theorem
  provers for first order logic and coherent logic so that it
  generates construction correctness proofs, that are both
  human-readable and formal (can be checked by interactive theorem
  provers such as Coq or Isabelle/HOL). These proofs currently rely on
  many high-level lemmas and our goal is to have them all formally
  shown from the basic axioms of geometry.
\end{abstract}


\section{Introduction}

Geometry construction problems are usually solved in four phases:

\begin{enumerate}
\item \emph{Analysis}: In this phase, the geometric figure
  to be constructed is analyzed. The specific constraints that apply to this 
  figure and the relationships between its elements are
  identified. By understanding the requirements and constraints, the steps required 
  to construct the desired figure can be determined.

\item \emph{Construction}: Once the problem is analyzed, the sequence
  of ruler and compass construction steps used to construct the figure
  is identified. 

\item \emph{Proof}: After the figure is constructed, it should be proved
  that it satisfies the properties and conditions given by the
  specification. Proofs in ruler and compass constructions often
  involve using geometric principles, such as the properties of
  angles, congruence, or similarity. A formal proof can be used to demonstrate 
  the validity of the construction and ensure that it meets
  the desired criteria.

\item \emph{Discussion}: The discussion phase involves reflection on
  the construction, its properties, and the relevant insights. It is
  often discussed under which condition does the solution exist and
  whether it is unique. Non-degeneracy conditions are also identified.
\end{enumerate}

In our previous work we have described our system ArgoTriCS that can
perform triangle constructions both in Euclidean
geometry~\cite{argotrics} and in absolute and hyperbolic
geometry~\cite{argotrics-hyperbolic}. Problems from the Wernick's
list~\cite{wernick-corpus} are analyzed and in Euclidean setting
ArgoTriCS manages to solve 66  out of 74 non-isomorphic
problems. Essentially it performs the problem analysis based on its
internal set of definitions and lemmas, and finds a series of
construction steps required to construct a triangle with a given set
of significant points (e.g., vertices, orthocenter, centroid, centers
of inscribed and circumscribed circles etc.). However it did not generate
classic, readable, synthetic construction proofs. In her PhD
thesis~\cite{marinkovic-phd}, Marinkovi\'c describes how theorem provers, based on algebraic methods such as Wu's method~\cite{Wu} and Gr\" obner basis method~\cite{groebner},
and semi-synthetic methods such as area method~\cite{area},
integrated within GLCL tool~\cite{gclc} and OpenGeoProver~\cite{GATPformalization},  
could be employed to check the construction
correctness. The problem with this approach is that generated proofs are not
human-readable. Since the main usage scenario of automated
construction solver is in education, it is vital that students
understand why some construction is correct. Therefore, obtaining
human-readable proofs is of a great importance.

In the current work, we describe how an automated system such as
ArgoTriCS can be combined with first-order logic and coherent logic 
provers so that each generated construction is accompanied by its 
human-readable proof of correctness. This is a work in progress, and 
we will describe our approach, prototype implementation, and preliminary 
results for a small set of selected problems.

\section{Examples}

\begin{example1}\label{ex:AHaO}
Consider constructing a triangle $ABC$ given its vertex $A$, altitude foot
$H_a$ and circumcenter $O$. ArgoTriCS finds the following
construction, illustrated in Figure \ref{fig:construction_0015}:

\begin{enumerate}
\item Construct the line $l_1 = AH_a$.
\item Construct the line $l_2$ such that it is perpendicular to the line
  $l_1$ and that it contains $H_a$.
\item Construct the circle $c$ centered at $O$ containing $A$.
\item Let $B$ and $C$ be the intersections of the line $l_2$ and the circle $c$.
\end{enumerate}

\begin{figures}[0.65]{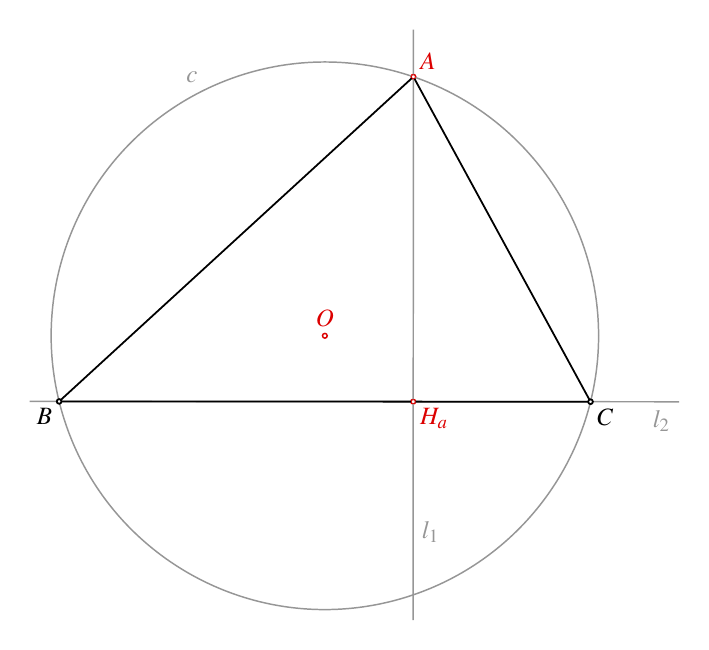}{Construction of the triangle $ABC$ given the points $A$, $O$, and $H_a$.}{fig:construction_0015}\end{figures}
\end{example1}

\begin{proof}
We need to show that $A$ is the vertex of the constructed
triangle $ABC$ (which is trivial), that $H_a$ is its altitude foot and
that $O$ is its circumcenter. This proof is rather straightforward.

By construction, the circle $c$ contains all three vertices $A$, $B$,
and $C$, so it must be the circumcircle of the triangle $ABC$ (since
the circumcircle of a triangle is unique). The $O$ is the center of $c$, so it must be
the circumcenter (since the center of any circle is unique).

By construction the line $l_2$ contains the vertices $B$ and $C$, so it
must be equal to the side $a$ of the triangle $ABC$ (since the triangle
side through the points $B$ and $C$ is unique). By construction the line $l_1$ contains
$A$ and is perpendicular to $l_2=a$, so it must be equal to the
altitude $h_a$ (since there is a unique altitude from the vertex $A$). Since by construction $H_a$ belongs both to
$l_2=a$ and $l_1 = h_a$ it must be the altitude foot $H_a$ (since it is
the unique intersection of $a$ and $h_a$).
\end{proof}

If we analyze the previous proof, we see that it essentially relies on several uniqueness lemmas and that it merely reverses the chain of
deduction steps used in the analysis phase.

In some cases, however, the proof is very different from the analysis.

\begin{example1}\label{ex:AOG}
Consider constructing a triangle $ABC$ given its vertex $A$, circumcenter
$O$ and centroid $G$. The construction that ArgoTriCS finds is the following (see Figure \ref{fig:construction_0031}):

\begin{enumerate}
\item Construct the point $P_1$ such that
  $\overrightarrow{AG} : \overrightarrow{AP_1} = 2:3$.
\item Construct the point $P_2$ such that
  $\overrightarrow{OG}:\overrightarrow{OP_2} = 1:3$.
\item Construct the line $l_1 = AP_2$.
\item Construct the line $l_2$ such that it is perpendicular to the line $l_1$
  and that it contains $P_1$.
\item Construct the circle $c$ centered at $O$ containing $A$.
\item Let $B$ and $C$ be the intersections of the line $l_2$ with the
  circle $c$.
\end{enumerate}

\begin{figures}[0.65]{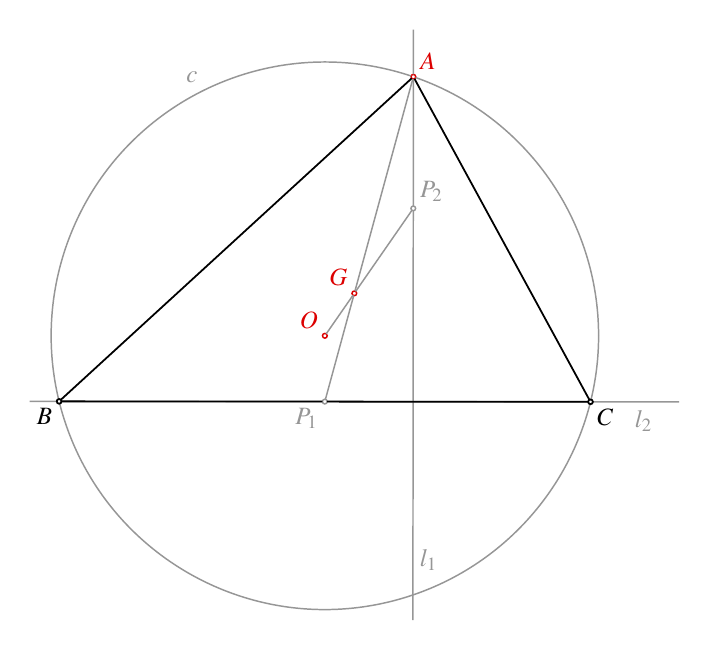}{Construction of the triangle $ABC$ given the points $A$, $O$, and $G$.}{fig:construction_0031}\end{figures}

Please note that there is a simpler solution to this construction problem, but we wanted 
to discuss this solution because the proof here is quite different from the construction.
\end{example1}

\begin{proof}
We need to prove that $A$ is the vertex of the triangle $ABC$ (which
is trivial), that $G$ is its centroid and that $O$ is its
circumcenter. The latter is very simple (similar to the previous
proof), since by construction all points $A$, $B$, and $C$ lie on the
circle $c$ centered at $O$.

The line $l_2$ is equal to the triangle side $a$, since it contains
the vertices $B$ and $C$ (and the triangle side through the points $B$ and $C$ is unique). By
construction $l_1$ contains $A$ and is perpendicular to $l_2 = a$, so
it must be equal to the altitude $h_a$ (since the altitude from vertex $A$ is unique).

Consider line $l_3 = OP_1$. We shall prove that it is parallel to the
line $l_1 = h_a$. Since by construction it holds that
$\overrightarrow{OG}:\overrightarrow{OP_2} = 1:3$, by the elementary
properties of vector ratio it also holds that
$\overrightarrow{OG}:\overrightarrow{GP_2} = 1:2$. Similarly, it holds
that $\overrightarrow{P_1G} : \overrightarrow{GA} = 1:2$. The angles
$OGP_1$ and $OGP_2$ are opposite and therefore congruent. Hence
triangles $OGP_1$ and $P_2GA$ are similar, and angles $OP_1G$ and
$GAP_2$ are always equal, so the lines $OP_1 = l_3$ and
$AP_2 = l_1 = h_a$ are parallel.

Since $h_a$ is perpendicular to $l_2 = a$, so must be $l_3 = OP_1$.
Therefore, the line $l_3$ must be the perpendicular bisector of the
segment $BC$ (since it is the unique line containing circumcenter $O$ that is
perpendicular to $a$). Consequently, the point $P_1$ must be equal to $M_a$ --
the midpoint of $BC$ (as it is the unique intersection of the segment with its
pependicular bisector). Finally, the point $G$ must
be the centroid of $ABC$ since the centroid is the unique point for
which it holds that $\overrightarrow{AG}:\overrightarrow{AM_a} = 2:3$.
\end{proof}

\section{Automation}

Our main goal is to obtain proofs such as the previous ones
automatically, using coherent logic provers.

\subsection{Problem Statement and Lemmas}

The first step would be to make
ArgoTriCS generate the problem statement, along with the construction
steps. For example, the problem statement for the first problem can be given as follows:
\begin{eqnarray*}
 && \textup{inc}(A, l_1) \wedge \textup{inc}(H_a', l_1) \wedge \\
 && \textup{perp}(l_2, l_1) \wedge \textup{inc}(H_a', l_2) \wedge \\
 && \textup{circle}(O', A, c) \wedge \\
 && \textup{inc}(B, l_2) \wedge \textup{inc}(C, l_2) \wedge \textup{inc\_c}(B, c) \wedge \textup{inc\_c}(C, c) \wedge B \neq C \Longrightarrow\\
 && H_a' = H_a \wedge O' = O
\end{eqnarray*}

The predicate $\textup{inc}(P, l)$ denotes that the point $P$ is
incident to the line $l$ i.e., $P \in l$, $\textup{inc\_c}(P, c)$
denotes that the point $P$ is incident to the circle $c$ i.e.,
$P \in c$, $\textup{circle}(O, P, c)$ denotes that $c$ is the
circle centered at the point $O$ passing through the point $P$, and
$\textup{perp}(l_1,l_2)$ denotes that lines $l_1$ and $l_2$ are perpendicular. The point
$O$ is the real circumcenter of the triangle $ABC$ (this is implicitly
given by the lemmas that are given to the prover along with the
problem statement), and $H_a$ is the real altitude foot. For
simplicity various non-degeneracy conditions are added to the problem
statement (e.g., the conditions $H_a' \neq A$, $A \neq B$, $A \neq C$,
etc.) before it is given to the automated theorem prover.

Along with the problem statement, automated prover is given a series
of carefully chosen lemmas, that are treated as axioms. Most of those
lemmas follow from the general geometric knowledge, but are
instantiated for the significant points, lines and circles of the
triangle $ABC$. Each significant object is denoted by a constant
(e.g., $bc$ for the side $BC$, $O$ for the circumcenter, $M_a$ for the
midpoint of $BC$, $h_a$ for the altitude from $A$, $H_a$ for its foot,
$c^\circ$ for the circumcircle etc.). Lemmas that encode properties of
those objects are used both in analysis (by the ArgoTriCS) and in
proofs (by automated theorem provers). Some of those lemmas are:
\begin{eqnarray*}
\textup{inc}(B, bc) &\wedge& \textup{inc}(C, bc) \\
\textup{inc}(A, h_a) &\wedge& \textup{perp}(h_a, bc) \\
\overrightarrow{AG}:\overrightarrow{AM_a} &=& 2:3 \\
\textup{inc\_c}(A, c^\circ) &\wedge& \textup{inc\_c}(B, c^\circ) \wedge \textup{inc\_c}(C, c^\circ)
\end{eqnarray*}
  
However, proofs require additional lemmas that guarantee uniqueness of
objects. For example:
\begin{eqnarray*}
(\forall l)(\textup{inc}(A, l) \wedge \textup{perp}(l, bc) &\Longrightarrow& l = h_a) \\
(\forall c)(\textup{inc\_c}(A, c) \wedge \textup{inc\_c}(B, c) \wedge \textup{inc\_c}(C, c) &\Longrightarrow& c = c^\circ)
\end{eqnarray*}

Notice that uniqueness lemmas are given in instantiated way, meaning that they hold for some specific objects.
This choice was made in order to follow the implementation of ArgoTriCS, where most
of the knowledge is given in an instantiated way. However, the uniqueness axioms
could be given also in more general way.

Some general lemmas about properties of basic geometric predicates are
also needed. For example:
\begin{eqnarray}
(\forall l_1,l_2)(\textup{perp}(l_1, l_2) &\Longrightarrow& \textup{perp}(l_2, l_1))\nonumber\\
(\forall P_1, P_2)(\exists l)(\textup{inc}(P_1, l) &\wedge& \textup{inc}(P_2, l))\nonumber
\end{eqnarray}

All those lemmas are formulated as axioms and the problem statement is
formulated as a conjecture in TPTP
format.\footnote{\url{https://www.tptp.org/}} That file is
then given to some automated theorem prover. In our experiments we
used Vampire~\cite{vampire} and Larus~\cite{larus}. Vampire is a very
efficient, award winning FOL theorem prover. Its main drawback is that
it cannot generate readable proofs. We also used Larus~\cite{larus}
that is a coherent-logic prover able to generate readable proofs and
also formal proofs that can be checked by interactive theorem provers
such as Isabelle/HOL or Coq.

Our second example uses the notion of ratio of vectors. However
neither Vampire nor Larus have a native support for arithmetic
calculations. Therefore we introduced separate predicates for ratios
that frequently occur in geometric constructions (e.g., $1:2$, $1:3$,
$2:3$) and added lemmas that connect those ratios. For example:
$$(\forall A, B, C)(\textup{ratio13}(A, B, A, C) \Longrightarrow \textup{ratio12}(A, B, B, C))$$

The proof uses a result that follows from triangle similarity. We
encoded this in the following lemma:
\begin{eqnarray*}
&&  (\forall A, M, B, X, Y, ax, by) \\
&&  (\textup{ratio21}(A, M, M, B) \wedge \textup{ratio21}(X, M, M, Y) \wedge \\
&&  \textup{line}(A, X, ax) \wedge \textup{line}(B, Y, by) \Longrightarrow \textup{para}(ax, by))
\end{eqnarray*}

Also, in Euclidean geometry there are clear connections between
parallel and perpendicular lines.
$$(\forall l1, l2, a)\; (\textup{perp}(l_1, a) \wedge \textup{para}(l_1, l_2) \Longrightarrow \textup{perp}(l_2, a))$$

\subsection{Using Automated Provers}

The conjecture of the construction problem considered in Example \ref{ex:AHaO} can be formulated in TPTP format in the following way:
\begin{verbatim}
fof(th_A_Ha_O, conjecture, ( ( inc(pA,ha1) & inc(pHa1,ha1) 
    & perp(ha1,a1) & inc(pHa1,a1) & inc_c(pA,cc1) & center(pOc1,cc1) 
    & inc_c(pB,cc1) & inc(pB,a1) & inc_c(pC,cc1) & inc(pC,a1) ) 
    => ( pHa = pHa1 &  pOc = pOc1 ) ) ).
\end{verbatim}
\noindent where \verb|pHa| and \verb|pOc| are defined by the axioms as the foot of the altitude from vertex $A$ to side $BC$ and circumcenter of triangle $ABC$, respectively.

Larus successfully proved given conjecture as two separate statements, one for each of the facts in the conclusion. Key fragment of generated
readable proof is given below (all used geometry axioms are listed, others are the ones implied by equality):

\vspace{10pt}

\noindent
  
{\bfseries Axioms:} 
\begin{enumerate}
\item bc\_unique : $\forall L \; ( inc(pB, L)\wedge inc(pC, L) \Rightarrow L = bc\;)$
\item haA : $\forall H \; ( perp(H, bc)\wedge inc(pA, H) \Rightarrow ha = H\;)$
\item pHa\_def : $\forall H1 \; ( inc(H1, ha)\wedge inc(H1, bc) \Rightarrow H1 = pHa\;)$
\item cc\_unique : $\forall C \; ( inc\_c(pA, C)\wedge inc\_c(pB, C)\wedge inc\_c(pC, C) \Rightarrow C = cc\;)$
\item center\_unique : $\forall C \; \forall C1 \; \forall C2 \; ( center(C1, C)\wedge center(C2, C) \Rightarrow C1 = C2\;)$
\end{enumerate}

\hrulefill

\begin{theorem1}
  th\_A\_Ha\_O0 :

  $inc(pA, ha1)$ $\wedge$ $inc(pHa1, ha1)$ $\wedge$ $perp(ha1, a1)$ $\wedge$ $inc(pHa1, a1)$ $\wedge$ $inc\_c(pA, cc1)$

  $\wedge$ $center(pOc1, cc1)$ $\wedge$ $inc\_c(pB, cc1)$ $\wedge$ $inc(pB, a1)$ $\wedge$ $inc\_c(pC, cc1)$ $\wedge$ $inc(pC, a1)$

  $\Rightarrow$ $pHa = pHa1\;$
\end{theorem1}

\newcounter{proofstepnum}
\setcounter{proofstepnum}{0}

\noindent{\em Proof:}
\vspace{5pt}

\proofstep{0}{$pHa = pHa$ ({\scriptsize by MP, using axiom eqnativeEqSub0; instantiation:  $A$ $\mapsto$  $pHa$,  $B$ $\mapsto$  $pHa$,  $X$ $\mapsto$  $pHa$}) }
\proofstep{0}{$a1 = bc$ ({\scriptsize by MP, from $inc(pB, a1)$, $inc(pC, a1)$ using axiom bc\_unique; instantiation:  $L$ $\mapsto$  $a1$}) }
\proofstep{0}{$perp(ha1, bc)$ ({\scriptsize by MP, from $perp(ha1, a1)$, $a1 = bc$ using axiom perpEqSub1; instantiation:  $A$ $\mapsto$  $ha1$,  $B$ $\mapsto$  $a1$,  $X$ $\mapsto$  $bc$}) }
\proofstep{0}{$ha = ha1$ ({\scriptsize by MP, from $perp(ha1, bc)$, $inc(pA, ha1)$ using axiom haA; instantiation:  $H$ $\mapsto$  $ha1$}) }
\proofstep{0}{$inc(pHa1, ha)$ ({\scriptsize by MP, from $inc(pHa1, ha1)$, $ha = ha1$ using axiom incEqSub1; instantiation:  $A$ $\mapsto$  $pHa1$,  $B$ $\mapsto$  $ha1$,  $X$ $\mapsto$  $ha$}) }
\proofstep{0}{$inc(pHa1, bc)$ ({\scriptsize by MP, from $inc(pHa1, a1)$, $a1 = bc$ using axiom incEqSub1; instantiation:  $A$ $\mapsto$  $pHa1$,  $B$ $\mapsto$  $a1$,  $X$ $\mapsto$  $bc$}) }
\proofstep{0}{$pHa1 = pHa$ ({\scriptsize by MP, from $inc(pHa1, ha)$, $inc(pHa1, bc)$ using axiom pHa\_def; instantiation:  $H1$ $\mapsto$  $pHa1$}) }
\proofstep{0}{$pHa = pHa1$ ({\scriptsize by MP, from $pHa1 = pHa$, $pHa = pHa$ using axiom eqnativeEqSub0; instantiation:  $A$ $\mapsto$  $pHa$,  $B$ $\mapsto$  $pHa1$,  $X$ $\mapsto$  $pHa$}) }
\proofstep{0}{Proved by assumption! ({\scriptsize by QEDas})}

\noindent
\begin{theorem1}
  th\_A\_Ha\_O1 :

  $inc(pA, ha1)$ $\wedge$ $inc(pHa1, ha1)$ $\wedge$ $perp(ha1, a1)$ $\wedge$ $inc(pHa1, a1)$ $\wedge$ $inc\_c(pA, cc1)$

  $\wedge$ $center(pOc1, cc1)$ $\wedge$ $inc\_c(pB, cc1)$ $\wedge$ $inc(pB, a1)$ $\wedge$ $inc\_c(pC, cc1)$ $\wedge$ $inc(pC, a1)$

  $\Rightarrow$ $pOc = pOc1\;$
\end{theorem1}

\setcounter{proofstepnum}{0}

\noindent{\em Proof:}
\vspace{5pt}

\proofstep{0}{$center(pOc, cc)$ ({\scriptsize by MP, using axiom centerEqSub1; instantiation:  $A$ $\mapsto$  $pOc$,  $B$ $\mapsto$  $cc$,  $X$ $\mapsto$  $cc$}) }
\proofstep{0}{$cc1 = cc$ ({\scriptsize by MP, from $inc\_c(pA, cc1)$, $inc\_c(pB, cc1)$, $inc\_c(pC, cc1)$ using axiom cc\_unique; instantiation:  $C$ $\mapsto$  $cc1$}) }
\proofstep{0}{$center(pOc1, cc)$ ({\scriptsize by MP, from $center(pOc1, cc1)$, $cc1 = cc$ using axiom centerEqSub1; instantiation:  $A$ $\mapsto$  $pOc1$,  $B$ $\mapsto$  $cc1$,  $X$ $\mapsto$  $cc$}) }
\proofstep{0}{$pOc = pOc1$ ({\scriptsize by MP, from $center(pOc, cc)$, $center(pOc1, cc)$ using axiom center\_unique; instantiation:  $C$ $\mapsto$  $cc$,  $C1$ $\mapsto$  $pOc$,  $C2$ $\mapsto$  $pOc1$}) }
\proofstep{0}{Proved by assumption! ({\scriptsize by QEDas})}

\vspace{10pt}

Correctness proof of the generated construction for the problem considered in Example \ref{ex:AOG} is given in Appendix.

\section{Results}

We considered the subset of problems from Wernick's corpus, over vertices of
the triangle, midpoints of triangle sides, feet of altitudes, centroid,
circumcenter and orthocenter of the triangle. It consists of 35 non-isomorphic
location triangle problems. For each of these problems, we tried to prove
the correctness of constructions found by ArgoTriCS using FOL prover Vampire and coherent logic
prover Larus. Vampire succesfully proved 31 of these problems, while Larus
proved 20 problems, and for remaining ones it could not prove it in given
timelimit.

\section{Conclusion}

Although this is a work-in-progress, we have managed to show
that this approach is plausible and can be used to automatically
obtain readable proofs of correctness for geometric constructions. This
is very important in the context of mathematical education, where
students need to know why a geometric statement holds. In our previous
work, we have described ArgoTriCS -- a system that is able to perform ruler
and compass construction steps for almost all solvable problems in the
Wernick's corpus~\cite{argotrics,wernick-corpus}. The main step in the
ArgoTriCS implementation was to formulate a good set of lemmas to be
used for analysing and finding the construction. This work shows
that an identified set of lemmas is not sufficient to generate
correctness proofs, and that the proof phase requires an additional set
of lemmas (mainly the lemmas that guarantee uniqueness, but also some
other equally important lemmas). However, once these lemmas are
identified, they can be passed to general-purpose theorem provers, which
can then generate fully synthetic proofs of correctness. Although the
coherent logic solvers we  have tested are not yet as powerful as the
FOL solvers such as Vampire, if they succeed in solving the given problem,
they provide us with human-readable proofs.

A very important issue is the correctness of the used lemmas. Indeed, if
some lemmas are incorrect (e.g., if a precondition or a
non-degeneracy condition is missing), a contradiction may arise and the
theorem could be proved from this contradiction. We examined all the
generated proofs, and all of them were correct. To be completely sure
that our lemmas are correct, we formalize them in Isabelle/HOL
and prove them using the axioms of geometry. Since Larus can
output Isabelle/HOL proofs, we will eventually have a system capable of
generating proofs of construction that are fully mechanically verified
starting from the axioms.

In the present work we have not considered degenerate cases and the existence
of constructed objects (we have simply assumed that everything is
non-degenerate and that all constructed objects exist). However, we
plan to pay more attention to this issue and extend our tools to
perform the final discussion phase where they would automatically 
identify the necessary non-degeneracy conditions.

Coherent logic prover, Larus, used in this research is currently unable 
to find all correctness proofs fully automatically. We have worked
around this by giving it hints in the form of lemmas. We plan to use 
other coherent logic provers, and we are in contact with the
Larus developers so that they can improve their prover using the
feedback they have received from our problems.

\appendix

\section{Appendix}

Larus cannot currently prove the whole theorem only if no guidance
is provided. Therefore, we first derive several lemmas and then use
those lemmas to prove the main theorem.
The first part of the conjecture is easily proved:

\vspace{10pt}

\noindent
  
{\bfseries Axioms:} 
\begin{enumerate}
\item cc\_unique : $\forall C \; ( inc\_c(pA, C)\wedge inc\_c(pB, C)\wedge inc\_c(pC, C) \Rightarrow C = cc\;)$
\item center\_unique : $\forall C \; \forall C1 \; \forall C2 \; ( center(C1, C)\wedge center(C2, C) \Rightarrow C1 = C2\;)$
\item bc\_unique : $\forall L \; ( inc(pB, L)\wedge inc(pC, L) \Rightarrow L = bc\;)$
\item haA : $\forall H \; ( perp(H, bc)\wedge inc(pA, H) \Rightarrow ha = H\;)$
\item inc\_line : $\forall P1 \; \forall P2 \; \forall L \; ( inc(P1, L)\wedge inc(P2, L)\wedge P1 \neq P2 \Rightarrow line(P1, P2, L)\;)$
\item ex\_line : $\forall P1 \; \forall P2 \; ( \exists L \; (line(P1, P2, L))\;)$
\item ratio21\_para : $\forall A \; \forall G \; \forall Ma \; \forall H \; \forall Oc \; \forall Lba \; \forall Lha \; ( ratio21(A, G, G, Ma)\wedge ratio21(H, G, G, Oc)\wedge line(Oc, Ma, Lba)\wedge line(A, H, Lha) \Rightarrow para(Lba, Lha)\;)$
\item perp\_para : $\forall Lba \; \forall Lha \; \forall A \; ( perp(Lha, A)\wedge para(Lba, Lha) \Rightarrow perp(Lba, A)\;)$
\item perp\_unique : $\forall P \; \forall L \; \forall L1 \; \forall L2 \; ( perp(L1, L)\wedge inc(P, L1)\wedge perp(L2, L)\wedge inc(P, L2) \Rightarrow L1 = L2\;)$
\item pMa\_is\_interect\_bisa\_bc : $\forall P \; ( inc(P, bc)\wedge inc(P, bisa) \Rightarrow P = pMa\;)$
\end{enumerate}

\hrulefill

\begin{theorem1}
  th\_A\_O\_G\_1:

  $ratio23(pA, pG1, pA, pMa1)$ $\wedge$ $ratio23(pH1, pG1, pH1, pOc1)$ $\wedge$ $inc(pA, ha1)$ $\wedge$ $inc(pH1, ha1)$ 

  $\wedge$ $inc(pMa1, a1)$ $\wedge$ $perp(a1, ha1)$ $\wedge$ $center(pOc1, cc1)$ $\wedge$ $inc\_c(pA, cc1)$ $\wedge$ $inc\_c(pB, cc1)$

  $\wedge$ $inc(pB, a1)$ $\wedge$ $inc\_c(pC, cc1)$ $\wedge$ $inc(pC, a1)$ $\wedge$ $pA \neq pH1$ $\Longrightarrow$ $pOc1 = pOc\;$
\end{theorem1}

\setcounter{proofstepnum}{0}

\noindent{\em Proof:}
\vspace{5pt}

\proofstep{0}{$cc1 = cc$ ({\scriptsize by MP, from $inc\_c(pA, cc1)$, $inc\_c(pB, cc1)$, $inc\_c(pC, cc1)$ using axiom cc\_unique; instantiation:  $C$ $\mapsto$  $cc1$}) }
\proofstep{0}{$center(pOc1, cc)$ ({\scriptsize by MP, from $center(pOc1, cc1)$, $cc1 = cc$ using axiom centerEqSub1; instantiation:  $A$ $\mapsto$  $pOc1$,  $B$ $\mapsto$  $cc1$,  $X$ $\mapsto$  $cc$}) }
\proofstep{0}{$pOc1 = pOc$ ({\scriptsize by MP, from $center(pOc1, cc)$ using axiom center\_unique; instantiation:  $C$ $\mapsto$  $cc$,  $C1$ $\mapsto$  $pOc1$,  $C2$ $\mapsto$  $pOc$}) }
\proofstep{0}{Proved by assumption! ({\scriptsize by QEDas})}

\vspace{5pt}
\noindent

Then, the facts \verb|a1 = bc| and \verb|ha1 = ha| can be derived:

\begin{lemma1}
  lm\_A\_O\_G\_2:

  $ratio23(pA, pG1, pA, pMa1)$ $\wedge$ $ratio23(pH1, pG1, pH1, pOc1)$ $\wedge$ $inc(pA, ha1)$ $\wedge$ $inc(pH1, ha1)$

  $\wedge$ $inc(pMa1, a1)$ $\wedge$ $perp(a1, ha1)$ $\wedge$ $center(pOc1, cc1)$ $\wedge$ $inc\_c(pA, cc1)$ $\wedge$ $inc\_c(pB, cc1)$

  $\wedge$ $inc(pB, a1)$ $\wedge$ $inc\_c(pC, cc1)$ $\wedge$ $inc(pC, a1)$ $\wedge$ $pA \neq pH1$ $\Longrightarrow$ $a1 = bc\;$
\end{lemma1}

\setcounter{proofstepnum}{0}

\noindent{\em Proof:}
\vspace{5pt}

\proofstep{0}{$a1 = bc$ ({\scriptsize by MP, from $inc(pB, a1)$, $inc(pC, a1)$ using axiom bc\_unique; instantiation:  $L$ $\mapsto$  $a1$}) }
\proofstep{0}{Proved by assumption! ({\scriptsize by QEDas})}

\vspace{5pt}
\noindent

\begin{lemma1}
  lm\_A\_O\_G\_3:

  $ratio23(pA, pG1, pA, pMa1)$ $\wedge$ $ratio23(pH1, pG1, pH1, pOc1)$ $\wedge$ $inc(pA, ha1)$ $\wedge$ $inc(pH1, ha1)$

  $\wedge$ $inc(pMa1, a1)$ $\wedge$ $perp(a1, ha1)$ $\wedge$ $center(pOc1, cc1)$ $\wedge$ $inc\_c(pA, cc1)$ $\wedge$ $inc\_c(pB, cc1)$

  $\wedge$ $inc(pB, a1)$ $\wedge$ $inc\_c(pC, cc1)$ $\wedge$ $inc(pC, a1)$  $\wedge$ $pA \neq pH1$ $\Longrightarrow ha1 = ha\;$
\end{lemma1}

\setcounter{proofstepnum}{0}

\noindent{\em Proof:}
\vspace{5pt}

\proofstep{0}{$a1 = bc$ ({\scriptsize by MP, from $inc(pB, a1)$, $inc(pC, a1)$ using axiom bc\_unique; instantiation:  $L$ $\mapsto$  $a1$}) }
\proofstep{0}{$perp(bc, ha1)$ ({\scriptsize by MP, from $perp(a1, ha1)$, $a1 = bc$ using axiom perpEqSub0; instantiation:  $A$ $\mapsto$  $a1$,  $B$ $\mapsto$  $ha1$,  $X$ $\mapsto$  $bc$}) }
\proofstep{0}{$ha = ha1$ ({\scriptsize by MP, from $perp(bc, ha1)$, $inc(pA, ha1)$ using axiom haA; instantiation:  $H$ $\mapsto$  $ha1$}) }
\proofstep{0}{$ha1 = ha$ ({\scriptsize by MP, from $ha = ha1$ using axiom eq\_sym; instantiation:  $A$ $\mapsto$  $ha$,  $B$ $\mapsto$  $ha1$}) }
\proofstep{0}{Proved by assumption! ({\scriptsize by QEDas})}

\vspace{5pt}
\noindent

Now the conclusions of these lemmas can be added to the set of
premises, and the next lemma can be proved:

\begin{lemma1}
  lm\_A\_O\_G\_4 :

  $ratio23(pA, pG1, pA, pMa1)$ $\wedge$ $ratio23(pH1, pG1, pH1, pOc1)$  $\wedge$ $inc(pA, ha1)$ $\wedge$ $inc(pH1, ha1)$

  $\wedge$ $inc(pMa1, a1)$ $\wedge$ $perp(a1, ha1)$ $\wedge$ $center(pOc1, cc1)$ $\wedge$ $inc\_c(pA, cc1)$ $\wedge$ $inc\_c(pB, cc1)$

  $\wedge$ $inc(pB, a1)$ $\wedge$ $inc\_c(pC, cc1)$ $\wedge$ $inc(pC, a1)$ $\wedge$ $pA \neq pH1$ $\wedge$ $pOc1 = pOc$ $\wedge$ $a1 = bc$ $\wedge$ $ha1 = ha$

  $\Longrightarrow$ $line(pOc1, pMa1, bisa)\;$
\end{lemma1}

\setcounter{proofstepnum}{0}

\noindent{\em Proof:}
\vspace{5pt}

\proofstep{0}{$inc(pOc1, bisa)$ ({\scriptsize by MP, from $pOc1 = pOc$ using axiom incEqSub0; instantiation:  $A$ $\mapsto$  $pOc$,  $B$ $\mapsto$  $bisa$,  $X$ $\mapsto$  $pOc1$}) }
\proofstep{0}{Let $w$ be such that $line(pOc1, pMa1, w)$ ({\scriptsize by MP, using axiom ex\_line; instantiation:  $P1$ $\mapsto$  $pOc1$,  $P2$ $\mapsto$  $pMa1$}) }
\proofstep{0}{$line(pA, pH1, ha1)$ ({\scriptsize by MP, from $inc(pA, ha1)$, $inc(pH1, ha1)$, $pA \neq pH1$ using axiom inc\_line; instantiation:  $P1$ $\mapsto$  $pA$,  $P2$ $\mapsto$  $pH1$,  $L$ $\mapsto$  $ha1$}) }
\proofstep{0}{$para(w, ha1)$ ({\scriptsize by MP, from $ratio23(pA, pG1, pA, pMa1)$, $ratio23(pH1, pG1, pH1, pOc1)$, $line(pOc1, pMa1, w)$, $line(pA, pH1, ha1)$ using axiom ratio21\_para; instantiation:  $A$ $\mapsto$  $pA$,  $G$ $\mapsto$  $pG1$,  $Ma$ $\mapsto$  $pMa1$,  $H$ $\mapsto$  $pH1$,  $Oc$ $\mapsto$  $pOc1$,  $Lba$ $\mapsto$  $w$,  $Lha$ $\mapsto$  $ha1$}) }
\proofstep{0}{$perp(ha1, bc)$ ({\scriptsize by MP, from $ha1 = ha$ using axiom perpEqSub0; instantiation:  $A$ $\mapsto$  $ha$,  $B$ $\mapsto$  $bc$,  $X$ $\mapsto$  $ha1$}) }
\proofstep{0}{$perp(w, bc)$ ({\scriptsize by MP, from $perp(ha1, bc)$, $para(w, ha1)$ using axiom perp\_para; instantiation:  $Lba$ $\mapsto$  $w$,  $Lha$ $\mapsto$  $ha1$,  $A$ $\mapsto$  $bc$}) }
\proofstep{0}{$w = bisa$ ({\scriptsize by MP, from $perp(w, bc)$, $line(pOc1, pMa1, w)$, $inc(pOc1, bisa)$ using axiom perp\_unique; instantiation:  $P$ $\mapsto$  $pOc1$,  $L$ $\mapsto$  $bc$,  $L1$ $\mapsto$  $w$,  $L2$ $\mapsto$  $bisa$}) }
\proofstep{0}{$line(pOc1, pMa1, bisa)$ ({\scriptsize by MP, from $line(pOc1, pMa1, w)$, $w = bisa$ using axiom lineEqSub2; instantiation:  $A$ $\mapsto$  $pOc1$,  $B$ $\mapsto$  $pMa1$,  $C$ $\mapsto$  $w$,  $X$ $\mapsto$  $bisa$}) }
\proofstep{0}{Proved by assumption! ({\scriptsize by QEDas})}

\vspace{5pt}
\noindent

Finally, with the conclusion of this lemma added to the premises, we
can prove the final statament:

\begin{theorem1}
  th\_A\_O\_G\_5:

  $ratio23(pA, pG1, pA, pMa1)$ $\wedge$ $ratio23(pH1, pG1, pH1, pOc1)$
  $\wedge$ $inc(pA, ha1)$ $\wedge$ $inc(pH1, ha1)$ $\wedge$
  $inc(pMa1, a1)$ $\wedge$ $perp(a1, ha1)$ $\wedge$ $center(pOc1, cc1)$
  $\wedge$ $inc\_c(pA, cc1)$ $\wedge$ $inc\_c(pB, cc1)$ $\wedge$
  $inc(pB, a1)$ $\wedge$ $inc\_c(pC, cc1)$ $\wedge$ $inc(pC, a1)$
  $\wedge$
  $pA \neq pH1$ $\wedge$ $pOc1 = pOc$ $\wedge$ $a1 = bc$ $\wedge$ $ha1 = ha$ $\wedge$ $pOc
  = pOc1$ $\wedge$ $line(pOc1, pMa1, bisa)$ $\Longrightarrow$ $pG = pG1\;$
\end{theorem1}

\setcounter{proofstepnum}{0}

\noindent{\em Proof:}
\vspace{5pt}

\proofstep{0}{$inc(pMa1, bc)$ ({\scriptsize by MP, from $inc(pMa1, a1)$, $a1 = bc$ using axiom incEqSub1; instantiation:  $A$ $\mapsto$  $pMa1$,  $B$ $\mapsto$  $a1$,  $X$ $\mapsto$  $bc$}) }
\proofstep{0}{$pMa1 = pMa$ ({\scriptsize by MP, from $inc(pMa1, bc)$, $line(pOc1, pMa1, bisa)$ using axiom pMa\_is\_interect\_bisa\_bc; instantiation:  $P$ $\mapsto$  $pMa1$}) }
\proofstep{0}{$ratio23(pA, pG1, pA, pMa)$ ({\scriptsize by MP, from $ratio23(pA, pG1, pA, pMa1)$, $pMa1 = pMa$ using axiom ratio23EqSub3; instantiation:  $A$ $\mapsto$  $pA$,  $B$ $\mapsto$  $pG1$,  $C$ $\mapsto$  $pA$,  $D$ $\mapsto$  $pMa1$,  $X$ $\mapsto$  $pMa$}) }
\proofstep{0}{$pG = pG1$ ({\scriptsize by MP, from $ratio23(pA, pG1, pA, pMa)$ using axiom ratio23\_Ma\_Gsat0; instantiation:  $X$ $\mapsto$  $pG1$}) }
\proofstep{0}{Proved by assumption! ({\scriptsize by QEDas})}

\vspace{5pt}
\noindent

\bibliographystyle{eptcs}
\bibliography{construction_proofs}

\end{document}